\documentclass[fleqn,10pt]{wlscirep}
\usepackage[utf8]{inputenc}
\usepackage[T1]{fontenc}

\usepackage{amssymb}
\usepackage{graphicx}
\usepackage{color}
\usepackage{hyperref}
\usepackage{url}
\hypersetup{colorlinks=true, citecolor=blue, urlcolor=blue, linkcolor=blue}
\usepackage{amssymb,amsfonts,amsmath,dsfont}
\usepackage{graphicx}
\usepackage{gensymb}
\usepackage{pifont}
\usepackage{bm}
\usepackage{appendix}
\usepackage{color}
\usepackage{accents}
\usepackage{mathtools}
\usepackage{multirow}

\begin{document}

\title{Control of magnetic states and spin interactions in bilayer CrCl$_{3}$ with strain and electric fields: An ab initio study}


\author[1,2,+]{Ali Ebrahimian}
\author[1,$\dagger$]{Anna Dyrda{\l}}
\author[3,*]{Alireza Qaiumzadeh}

\affil[1]{Department of Mesoscopic Physics, ISQI, Faculty of Physics, Adam Mickiewicz University, ul. Uniwersytetu Poznanskiego 2, 61-614 Poznan, Poland}
\affil[2]{School of Physics, Institute for Research in Fundamental Sciences (IPM),Tehran 19395-5531, Iran}
\affil[3]{Center for Quantum Spintronics, Department of Physics, Norwegian University of Science and Technology, NO-7491 Trondheim, Norway}

\affil[+]{aliebrahimian@ipm.ir}
\affil[$\dagger$]{adyrdal@amu.edu.pl}
\affil[*]{alireza.qaiumzadeh@ntnu.no}

\begin{abstract}
Using ab initio density functional theory (DFT), we demonstrated the possibility of controlling the magnetic ground-state properties of bilayer CrCl$_{3}$ by means of mechanical strains and electric fields. In principle, we investigated the influence of these two fields on parameters describing the spin Hamiltonian of the system. The obtained results show that biaxial strains change the magnetic ground state between ferromagnetic (FM) and antiferromagnetic (AFM) phases. The mechanical strain also affects the direction and amplitude of the magnetic anisotropy energy (MAE). Importantly, the direction and amplitude of the Dzyaloshinskii-Moriya vectors are also highly tunable under external strain and electric fields. The competition between nearest neighbor interaction, MAE, and Dzyaloshinskii-Moriya interactions can lead to the stabilization of various exotic spin textures and novel magnetic excitations. The high tunability of magnetic properties by external fields makes bilayer CrCl$_{3}$ a promising candidate for application in the emerging field of two-dimensional quantum spintronics and magnonics.

\end{abstract}

\maketitle

\section{Introduction}\label{sec:intro}
\begin{figure*}[t]
	\centering
\vspace{-1.5cm}
	\includegraphics[width=0.8\textwidth]{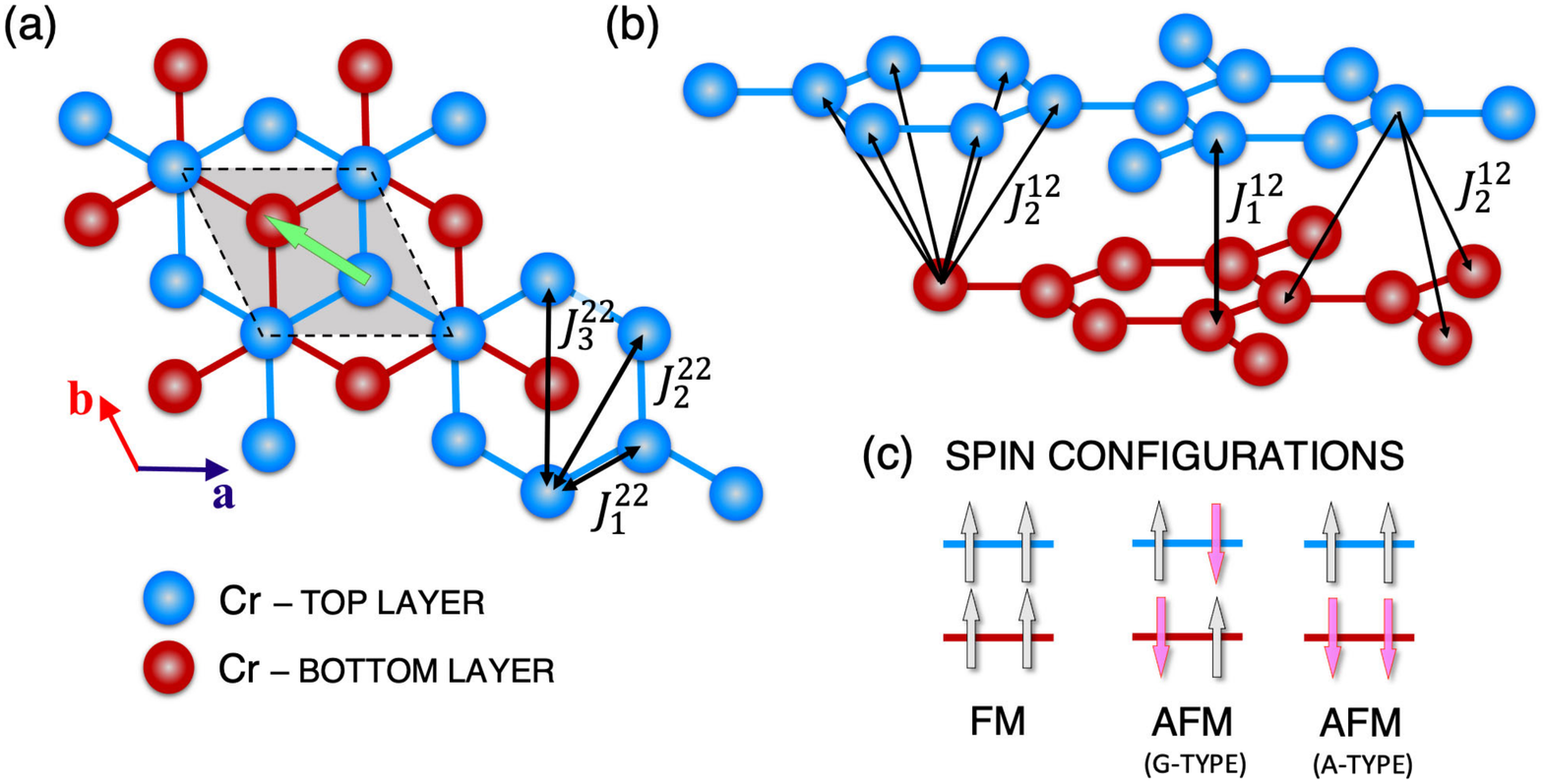}\\
\vspace{-1.5cm}	
\caption{(Color online) (a), (b) Arrangement of Cr atoms in the lattice of the CrCl$_{3}$ bilayer in the LT phase. The inter- and intralayer magnetic interactions between Cr atoms are indicated. The green arrow indicates the direction of lateral shift between the top and bottom layers. (c) Schematic plot of three different spin configurations of bilayer CrCl$_{3}$.
		\label{fig:1}}
\end{figure*}

The discovery of long-range magnetic order at finite temperature in atomically thin layers and the high tunability of their magnetic and electronic states represents a new route toward the next generation of ultrafast and energy-efficient spintronic-based nanodevices. Two-dimensional (2D) magnetic systems provide an ideal platform for investigating exotic properties arising from the interplay among different competing electronic and magnetic phenomena \cite{1,2,3,4,5,6,AlirezaUltrafast,AlirezaCrI3,Banasree}. The long-range magnetic order in these 2D layers is usually stabilized at finite temperature via intrinsic magnetic anisotropies and dipolar interactions \cite{PhysRevLett.77.386,fridman2002stabilization,girovsky2017long}.
Transition chromium trihalides (CrX$_{3}$, X= Cl, Br, I) represent a family of layered magnetic van der Waals (vdW) materials with potential applications in optoelectronics \cite{8} and quantum spintronics \cite{9}. Bulk CrX$_{3}$ is composed of Cr atoms arranged in hexagonal lattice layers, and each hexagonal layer is sandwiched between two halide planes, where Cr$^{3+}$ cations are octahedrally coordinated by six halide X$^{-}$ anions. The layers are bonded via weak vdW interactions. Bulk CrI$_{3}$ and CrBr$_{3}$ are ferromagnetic (FM) below their Curie temperatures, with magnetic moments aligned perpendicular to the Cr planes, while bulk CrCl$_{3}$ is antiferromagnetic (AFM), with a small in-plane anisotropy below its N{\'e}el temperature of 14 K \cite{10}. The most stable stacking arrangement of bulk CrX$_{3}$ at low temperature (LT) is a rhombohedral structure with R$\bar{3}$ space group symmetry, while at high temperature (HT), monoclinic layer stacking with C2/m space group symmetry is the stable crystalline phase \cite{11}.

The efficient control of spin interactions and magnetic phases in 2D systems represents important issues from technological and scientific points of view. Modifications of spin interactions cannot only tune the critical temperature of these materials by tuning their magnetic anisotropies but can also lead to the modification of magnetic states \cite{AlirezaUltrafast, Banasree}, emerging exotic magnetic textures \cite{song2021direct,yao2021noncollinear}, and magnonic phenomena \cite{AlirezaDWDMI} on demand.

The investigation of the magnetic properties of vdW materials such as CrX$_{3}$ shows that the spin interaction strengths between intra- and interlayer interactions are crucially different \cite{20} In fact, compared to intralayer spin interactions, interlayer spin interactions are relatively weak and become negligible beyond the nearest neighbor atomic layers. Therefore, tuning interlayer interactions should be possible by external fields. On the other hand, since the in-plane Cr-X-Cr angle ($\approx 95^{\circ}$) in these materials is close to the critical Goodenough-Kanamori-Anderson angle of $90^{\circ}$, the FM superexchange interaction between Cr atoms can be externally tuned and form AFM interactions \cite{14}.
Indeed, it has been shown that the magnetic properties of 2D vdW layers can be controlled by applying strain and electric fields \cite{Song_2019,Li_2019,12,13,AlirezaCrI3,Banasree,PhysRevLett.127.037204,Pizzochero_2020}.

Monolayer and bilayer of transition chromium trihalides (CrI$_{3}$), with a hexagonal magnetic lattice structure,
were respectively among the first 2D FM and AFM systems discovered
that revealed long-range magnetic orders with a critical transition temperature of about 45 K and strong out-of-plane magnetic anisotropy that opens a gap in the magnon dispersion at the $\Gamma$-point of the magnetic Brillouin zone (BZ) \cite{7,12, pizzochero2020magnetic, AlirezaCrI3}. This magnonic gap is crucial for existence of long-range magnetic order at finite temperature in 2D systems. 
Magnon spectrum in a hexagonal lattice with FM order has a Dirac-like dispersion at K points of the magnetic BZ \cite{PhysRevB.94.075401}. Recent experimental studies show that in monolayer FM CrI$_{3}$ there is a finite gap at these K points. The origin of this gap is still unclear and under intense debate \cite{olsen2021unified}. A few proposals show that this gap has a topological nature arising from either Dzyaloshinskii-Moriya or Kitaev interaction \cite{lee2020fundamental,chen2018topological,jaeschke2021theory}. Other theories relate this gap to electron correlations and hence a nontopological origin \cite{ke2021electron}. On the other hand,  bilayer of CrI$_{3}$ is an A-type AFM system, in which the intralayer exchange interaction is FM but the interlayer exchange interaction is AFM \cite{jiang2018electric,morell2019control}. Voltage and strain-controlled switching between AFM and FM phases as well as tuning of the electronic band gap have been reported for this structure \cite{14,15,16,17,18,leon2020strain,Song_2019,Li_2019,12,13}.

Although there are many theoretical and experimental studies on monolayer and bulk CrX$_{3}$ materials, there are few studies on their multilayer structures. Bilayers and trilayers of these materials have greater potential for applications due to their high tunability and functionality. The weak spin interactions between adjacent layers and their magnetic ground states have not been accurately explored, which results in a poor understanding of magnetic ordering in CrX$_{3}$ bilayers.
There is some discrepancy between first-principles calculations and experimental measurements regarding the layer stacking and magnetic phases in bilayer CrX$_{3}$ \cite{19, 20}.
Furthermore, due to the small energy difference between FM and AFM phases, in the HT phase, different ab initio calculations of magnetic states are not always compatible with each other, and the results are sensitive to the implemented methods and initial parameters \cite{20, 21, 22}.

In this paper, we investigate the magnetic properties of bilayer CrCl$_{3}$ in the presence of strain and electric fields. Monolayer CrCl$_{3}$ is a candidate for 2D-XY ferromagnetism and thus the Berezinskii-Kosterlitz-Thouless (BKT) phase transition\cite{bed}. It has been shown that the magnetic anisotropy and magnetic state of monolayer CrCl$_{3}$ can be tuned by strain and electric fields \cite{PhysRevLett.127.037204}.
To the best of our knowledge, no systematic investigations have been conducted on the control of spin interactions in bilayer CrCl$_{3}$. Therefore, we present a comprehensive study on the magnetic properties of bilayer CrCl$_{3}$ in the presence of biaxial strain and electric fields by means of ab initio calculations and magnetic force theory (MFT). We investigate how biaxial strain and electric fields tune the intra- and interlayer exchange interactions, Dzyaloshinskii-Moriya interactions (DMIs), and magnetic anisotropy.

The paper is organized as follows. In Sec.~\ref{sec:theory}, we briefly introduce our simulation methods and our minimal model for the spin Hamiltonian. The effects of strain and electric fields on the electronic and magnetic properties of bilayer CrCl$_{3}$ are presented in Sec.~\ref{sec:level3}. We conclude the paper in Sec.~\ref{conclusion}.

\section{Spin Hamiltonian of Bilayer CrCl$_{3}$ and Computational Methods}
\label{sec:theory}

A schematic plot of the bilayer CrCl${_3}$ is shown in Fig.~\ref{fig:1}. In the unit cell of monolayer CrCl$_{3}$, there are two magnetic Cr atoms, and each Cr atom is surrounded by six I atoms in accordance with the octet rule. The two adjacent layers are coupled by weak vdW forces. The basis vector $\bf{a}$ ($\bf{b}$) is along the zigzag (armchair) direction of the honeycomb lattice of the magnetic Cr atoms (see Fig.~\ref{fig:1}). A monolayer of CrCl$_{3}$ has three atomic sublayers. In the octahedral environment, electric crystal fields split the $d$ orbitals of the Cr atoms into a set of triply degenerate orbitals $t_{2g}$ (with lower energy) and doubly degenerate orbitals $e_{g}$ (with higher energy). Each of the three $t_{2g}$ orbitals is occupied by one electron to minimize both the orbital energy and Coulomb interaction energy and produce an atomic magnetic moment of 3 $\mu_B$. Our ab initio calculations show a magnetic moment of 3.17 $\mu_B$ for Cr atoms and an induced magnetic moment of -0.05 $\mu_B$ on the Cl atoms, which is in agreement with experiments \cite{32}. The CrCl$_{3}$ bilayer can be crystallized in rhombohedral structure with an R$\bar{3}$ space group symmetry at LT and monoclinic structure with the C2/m space group symmetry at HT. The total energy calculation shows that the LT phase of CrCl$_{3}$ is more stable and that its energy is lower than that of the HT phase, which is in a metastable state. Therefore, in this study, we consider only the magnetic properties of the LT phase. In the LT phase, the top layer is laterally shifted by $\left[-1/3, 1/3\right]$ in fractional coordinates with respect to the bottom layer (Fig.~\ref{fig:1}(a)).

Ab initio density functional theory (DFT) calculations were performed using the QUANTUM ESPRESSO package \cite{25}. In all calculations, we employed the Perdew-Burke-Ernzerhof (PBE) \cite{26} flavor for the generalized gradient exchange-correlation functional. The BZ was sampled by a $12\times 12\times 1$ k-point grid mesh, and a plane-wave cutoff energy of 80 Ry was considered. To avoid any interactions between the plane images, a 25 {\AA} vacuum was applied along the z-axis. The total ground-state energy converged to within an accuracy of $10^{-10}$ eV. Furthermore, the lattice parameters and atomic positions were optimized until the maximum force on each atom was less than $10^{-3}$ eV/{\AA}.

The vdW Grimme-D2 correction \cite{29} was used to consider the interaction between adjacent layers. Since the onsite Hubbard interaction is essential to finding the true ground state in 2D systems, we used the DFT+U method \cite{30}. In fact, as the band structure near the Fermi level is mostly composed of the localized $d$ orbitals of Cr atoms, we used the DFT+U method to take into account the effect of strong electron correlations. Considering the onsite Coulomb interaction of Cr-3$d$ orbitals, U = 3 eV in all calculations.
The selected value of U in our calculations is similar to an earlier choice in the previous investigation of the CrCl${_3}$ bilayer \cite{32} and is around the first-principles-derived values \cite{jang2019}. However, it should be noted that our calculation results do not depend on Hubbard U values and remain unchanged under variation in U. The robustness of the results against different U values is shown in the Supplementary Information.

Investigating the magnetic ground state, we considered three different magnetic configurations
(i) FM configuration, with all magnetic moments initialized in the same direction, (ii) A-type AFM configuration, with FM order at each layer and AFM coupling between two adjacent layers, and (iii) G-type AFM state, with magnetic moments coupled antiferromagnetically at each layer and between layers (see Fig.~\ref{fig:1} (c)).

To compute the single ion magnetic anisotropy energy (MAE), the total energy is computed by means of fully relativistic self-consistent-field DFT calculations incorporating spin-orbit coupling (SOC) and noncollinear spin-polarization effects.
The single ion MAE is defined as the difference between total energies corresponding to the magnetization orientation in-plane and out-of-plane, ${\mathrm{MAE}=E_{\scriptscriptstyle{\mathrm{IN}}}-E_{\scriptscriptstyle{\mathrm{OUT}}}}$, and computed within MFT \cite{27, 28, 31}. Therefore, a negative (positive) value of MAE indicates a uniaxial hard-axis (easy-axis) magnetic anisotropy.

\begin{figure}[t]
	\centering
	\includegraphics[width=0.99\linewidth]{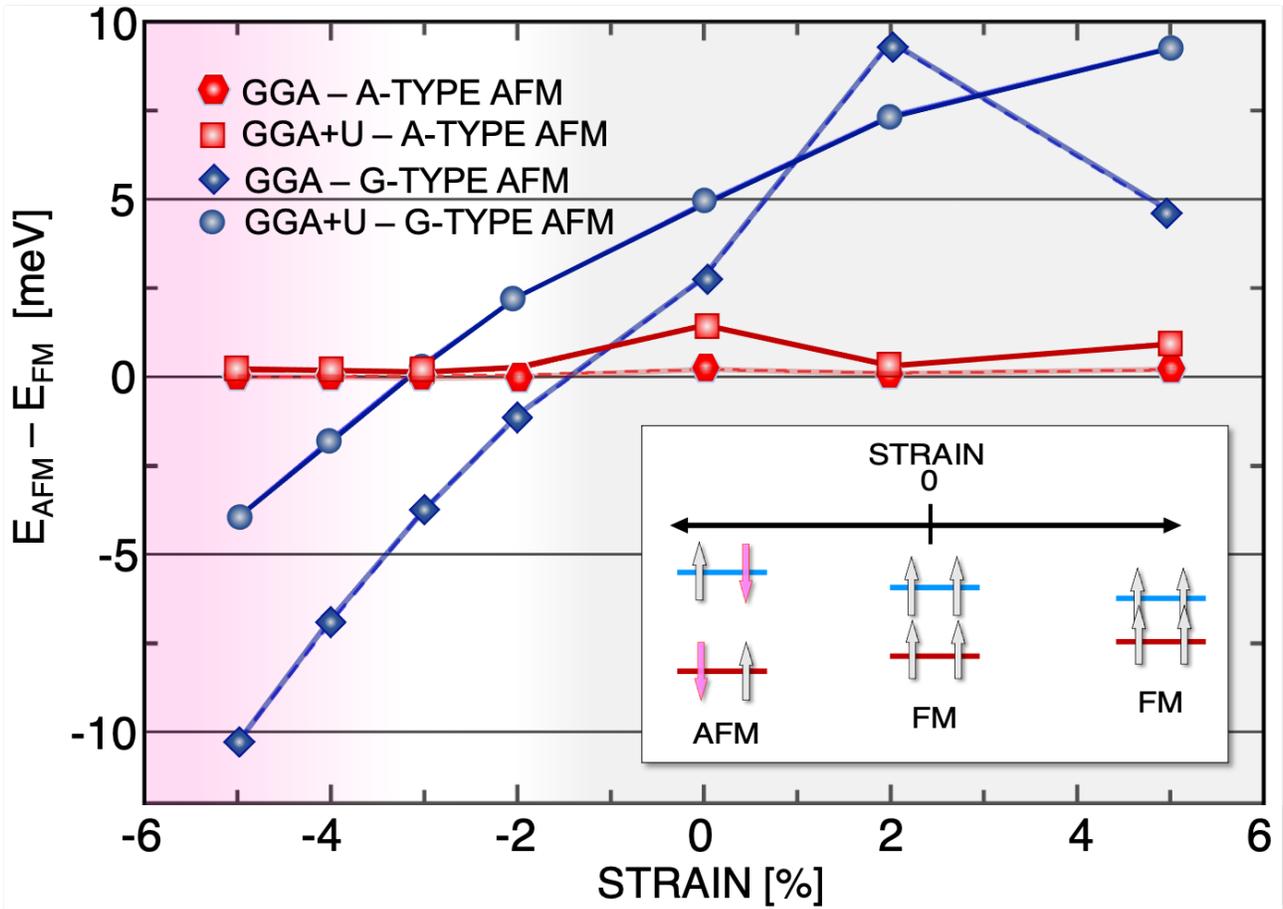}\\
	\caption{(Color online) The energy difference between the FM and AFM phases for CrCl$_{3}$ as a function of strain. The results of GGA+U (GGA) calculation indicate that a magnetic phase transition from FM to G-type AFM can occur under a compressive strain larger than 3(1)$\%$
		\label{fig:Fig2}}
\end{figure}

\begin{figure}[t]
	\centering
	\includegraphics[width=0.99\linewidth]{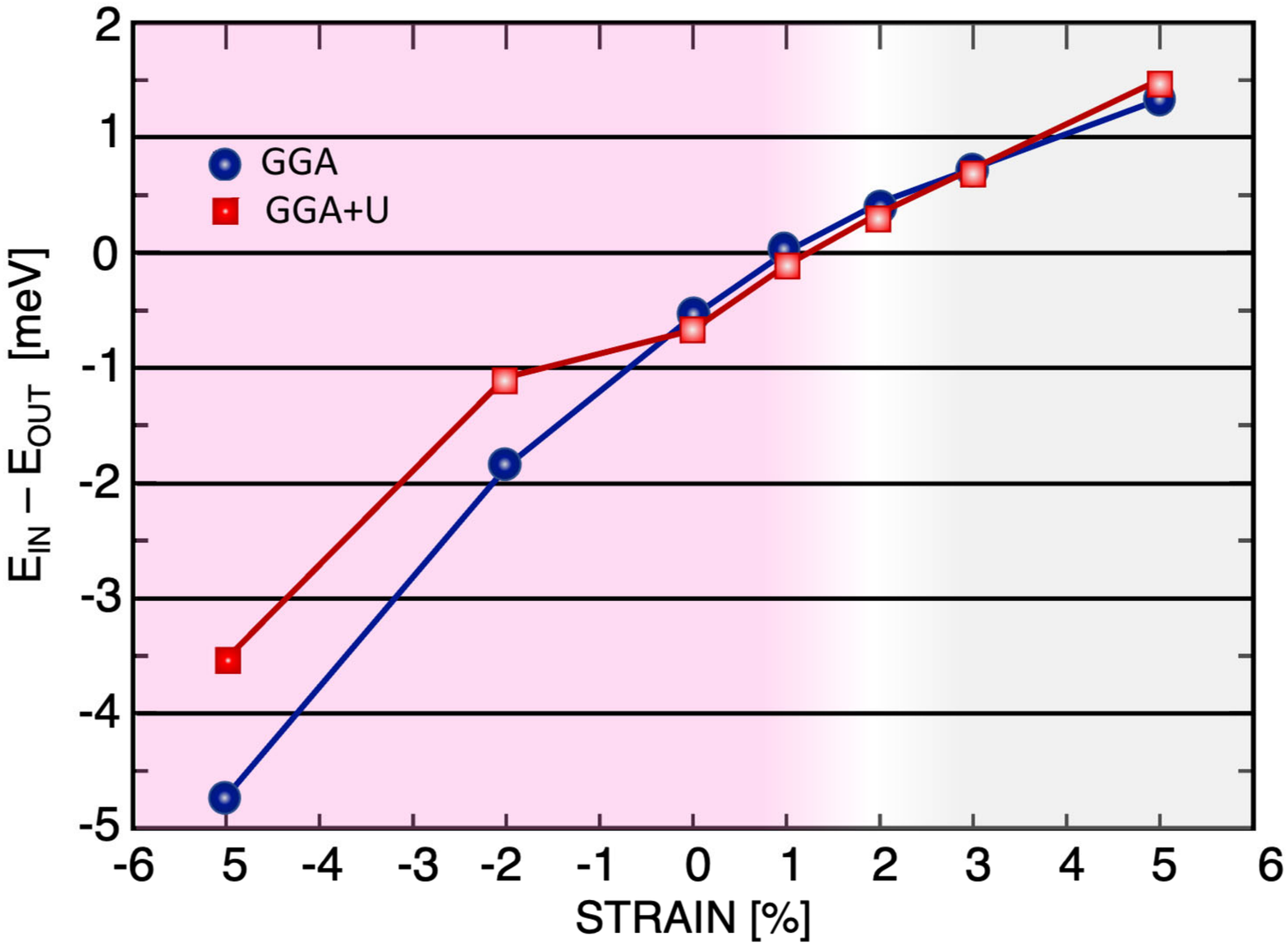}\\
	\caption{(Color online) Change in MAE of CrCl$_{3}$ bilayer with respect to the applied strain, within GGA and GGA+U approximations. A tensile strain equal to +2$\%$ induces a phase transition to an out-of-plane easy axis in this bilayer.
		\label{fig:D2}}
\end{figure}

To extract the spin-spin interactions in the bilayer CrCl$_{3}$ system, we adopt a minimal model spin Hamiltonian for $n$ magnetic moments arranged in hexagonal lattices within the XY plane of
\begin{align}
\label{eq:SpinH}
H =&- \sum^2_{l,l'=1} \sum^{n}_{\{i \neq j\}=1 } \left( {J}^{ll'}_{ij} {S}^{l}_{i}\cdot{S}^{l'}_{j} + {{D}}^{ll'}_{ij}\cdot({S}^{l}_{i} \times {S}^{l'}_{j})\right) 
 - \sum_{ l} \sum_{ i} {K}_{i} ({S}^{l}_{i}\cdot \hat{{z}})^{2},
\end{align}
where ${S}^{l}_{i}$ denotes the magnetic moment of a Cr atom in layer $l$ and at site $i$, $J^{l,l'}_{ij}$ represents the intra- ($l=l'$) and inter- ($l \neq l'$) layer symmetric Heisenberg exchange coupling between different magnetic moments at sites of $i$ and $j$, ${D}^{ll'}_{ij}$ denotes the intra- ($l=l'$) and inter- ($l \neq l'$) layer DMI vector, and $K_i >0 (<0)$ is the single ion easy-axis (hard-axis) magnetic anisotropy energy along the z-direction.
$J^{ll'}_{ij}>0$ indicates FM coupling, while $J^{ll'}_{ij}<0$ indicates AFM coupling. The direction of the DMI vector is dictated by the symmetry of the magnetic crystal, whereas its amplitude is proportional to the SOC strength \cite{AlirezaDMI1,AlirezaDMI2,AlirezaDMI3,AlirezaUltrafast}. The strength and sign of magnetic anisotropy are also dependent on SOC.

To obtain the spin-spin interaction parameters, presented in the spin Hamiltonian model~(\ref{eq:SpinH}), we use a Green’s function method with the local rigid spin rotation, treated as a perturbation \cite{31}. In fact, the rigid spin-rotation perturbation is done within the single-particle Green’s function formalism to perturb a localized spin in the DFT electronic model. Next, spin-spin interaction parameters are mapped to the electron expressions \cite{LIECHTENSTEIN198765}. Finally, the different parameters can be decomposed as the sum of the contributions from orbital pairs in two atoms \cite{LIECHTENSTEIN198765}.
In the next section, we compute different spin-spin interaction parameters in the spin Hamiltonian model~(\ref{eq:SpinH}). 

\begin{figure*}
\centering
\includegraphics[width=0.85 \textwidth]{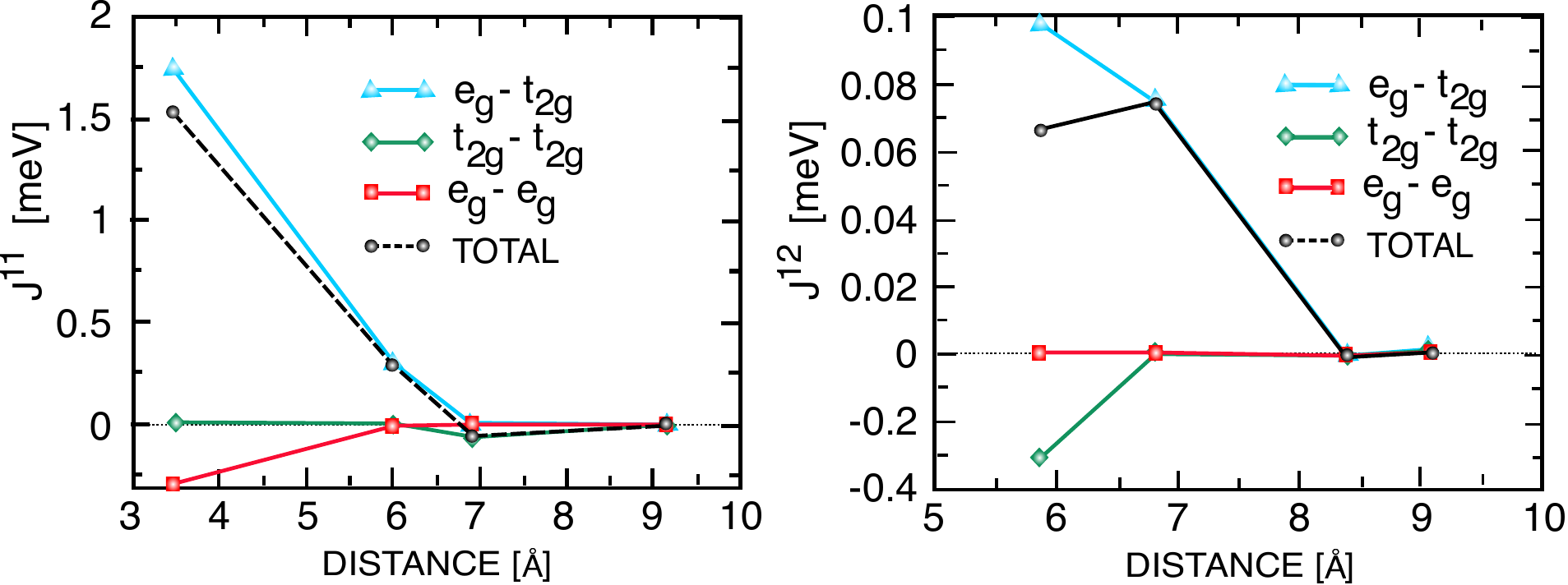} \\
\caption{(Color online) Total and orbitally resolved intralayer (left) and interlayer (right) exchange parameters of bilayer CrCl$_{3}$.
 \label{fig:E}}
\end{figure*}

\section{Numerical Results and Discussion}\label{sec:level3}

\subsection{Pristine Bilayer CrCl3}
First, we study the magnetic properties of a pristine CrCl$_{3}$ bilayer in the absence of any external field.
To find the magnetic ground state, we compute the energy differences between the FM state and two AFM states (types A and G; see Fig.~\ref{fig:1}(c)), $E_{AFM} - E_{FM}$. From ab initio calculations, it follows that the magnetic ground state of a CrCl$_{3}$ bilayer is an FM state~\cite{32}; see Fig.~\ref{fig:Fig2}. The FM ground state is robust against different values of the onsite Coulomb U. For the optimized structure of the CrCl$_{3}$ bilayer, we find a lattice constant of $a_{0}=5.99 {\AA}$, a Cr-Cl bond length of ${L_{Cr-Cl}=2.35\,\AA}$ and a bond angle of ${\alpha_{{Cr}-{Cl}-{Cr}}=94.55^{\circ}}$ (see the Supplementary Information).

To compute the single ion MAE, the relativistic SOC has been implemented in ab initio calculations. The MAE of the CrCl$_{3}$ bilayer is determined by calculating the energy difference between the in-plane and out-of-plane magnetic configurations; MAE=$E_{\scriptscriptstyle{{IN}}} - E_{\scriptscriptstyle{{OUT}}}$. Fig. \ref{fig:D2} shows that the MAE, and thus the magnetic anisotropy, $K_i$, of unstrained bilayer CrCl$_{3}$ are negative, indicating that the magnetic moments of the Cr atoms are in plane \cite{32}.

\begin{figure*}[t]
\centering
\includegraphics[width=0.9 \textwidth]{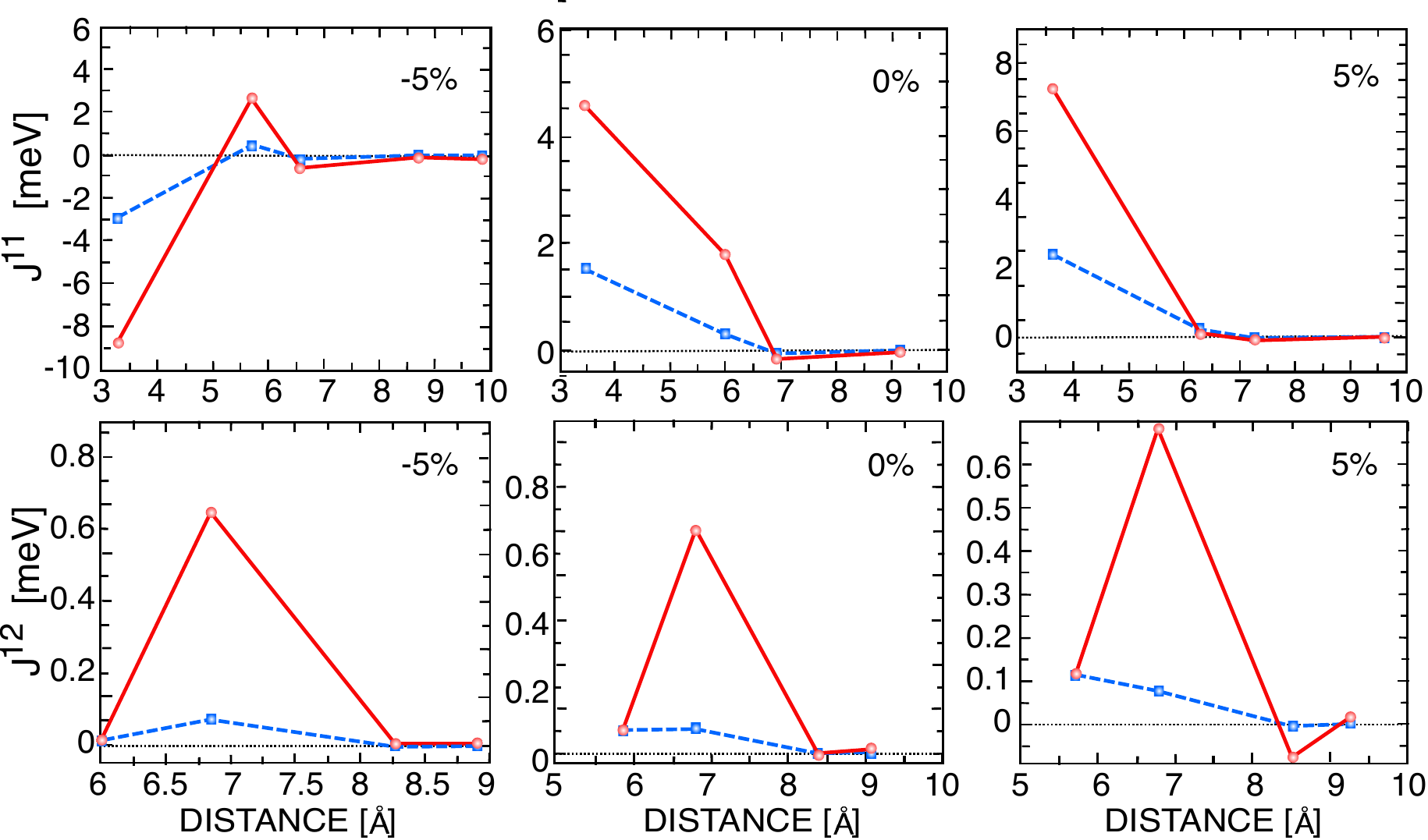} \\
\caption{(Color online) Total intralayer (top panel) and interlayer (bottom panel) exchange parameters (dashed line) summed over all neighboring pairs of bilayer CrCl$_{3}$ (solid line) for unstrained (middle), -5$\%$ strained (left) and +5$\%$ strained (right) structures. A positive value indicates FM contribution, and a negative value indicates AFM contribution.
\label{fig:F}}
\end{figure*}

Next, we calculate the total and orbital decomposed intra- and interlayer Heisenberg exchange interactions. Fig.~\ref{fig:E} shows that the intralayer magnetic interactions between Cr atoms are FM (consistent with the total energy analysis), with dominant values of 1.52 and 0.29 meV for the nearest neighbor (NN) and next nearest neighbor (NNN) interactions, respectively.
The NN intralayer exchange interactions between Cr atoms are mediated by $e_{g}$-$t_{2g}$ FM interactions and $e_{g}$-$e_{g}$ AFM interactions, while the NNN intralayer exchange interactions result mainly from $e_{g}$-$t_{2g}$ FM channels. The 3rd NN interaction is AFM and contributed mostly from $t_{2g}$-$t_{2g}$ coupling.
Due to the longer distance between Cr atoms, the intralayer interactions between neighbors further away are negligible. Within the same layer, the Cr atom has three NN, six NNN and three 3rd NN couplings. Therefore, the sum over all intralayer neighboring pairs shows that the NN exchange interaction makes the main contribution to intralayer FM coupling (Fig.~\ref{fig:F}).
\begin{figure*}[t]
\centering
\includegraphics[width=0.96 \textwidth]{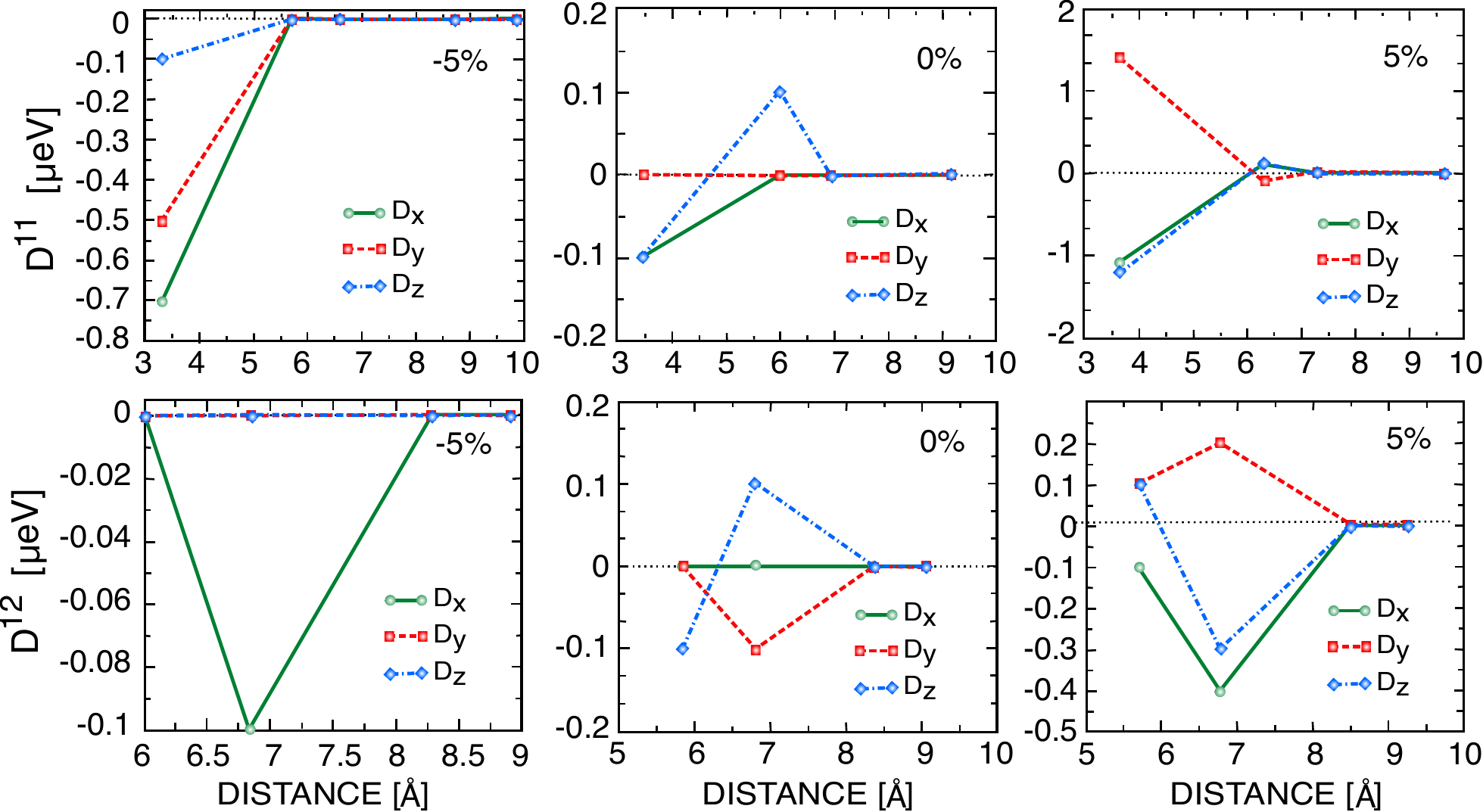} \\
\caption{(Color online) Intralayer (top panel) and interlayer (bottom panel) DMIs of bilayer CrCl$_{3}$ for unstrained (middle), -5$\%$ strained (left) and +5$\%$ strained (right) structures.
\label{fig:dm}}
\end{figure*}

The interlayer exchange interactions between two adjacent CrCl$_{3}$ layers are computed in a similar way as intralayer interactions. The LT stacking allows one NN, nine NNN, and twelve 3rd NN couplings with larger distances than the corresponding couplings in the same layer, which results in weak interlayer exchange couplings (Fig.~\ref{fig:E}). The NN and NNN exchange couplings are FM, where the dominant contribution comes from $e_{g}$-$t_{2g}$ couplings. Although the AFM $t_{2g}$-$t_{2g}$ channels make a noticeable contribution to first neighbor interactions, they become negligible for larger distances. Therefore, our ab initio calculations show FM interlayer coupling upon LT stacking of the CrCl$_{3}$ bilayer.

\begin{figure*}[t]
\centering
\includegraphics[width=0.75 \textwidth]{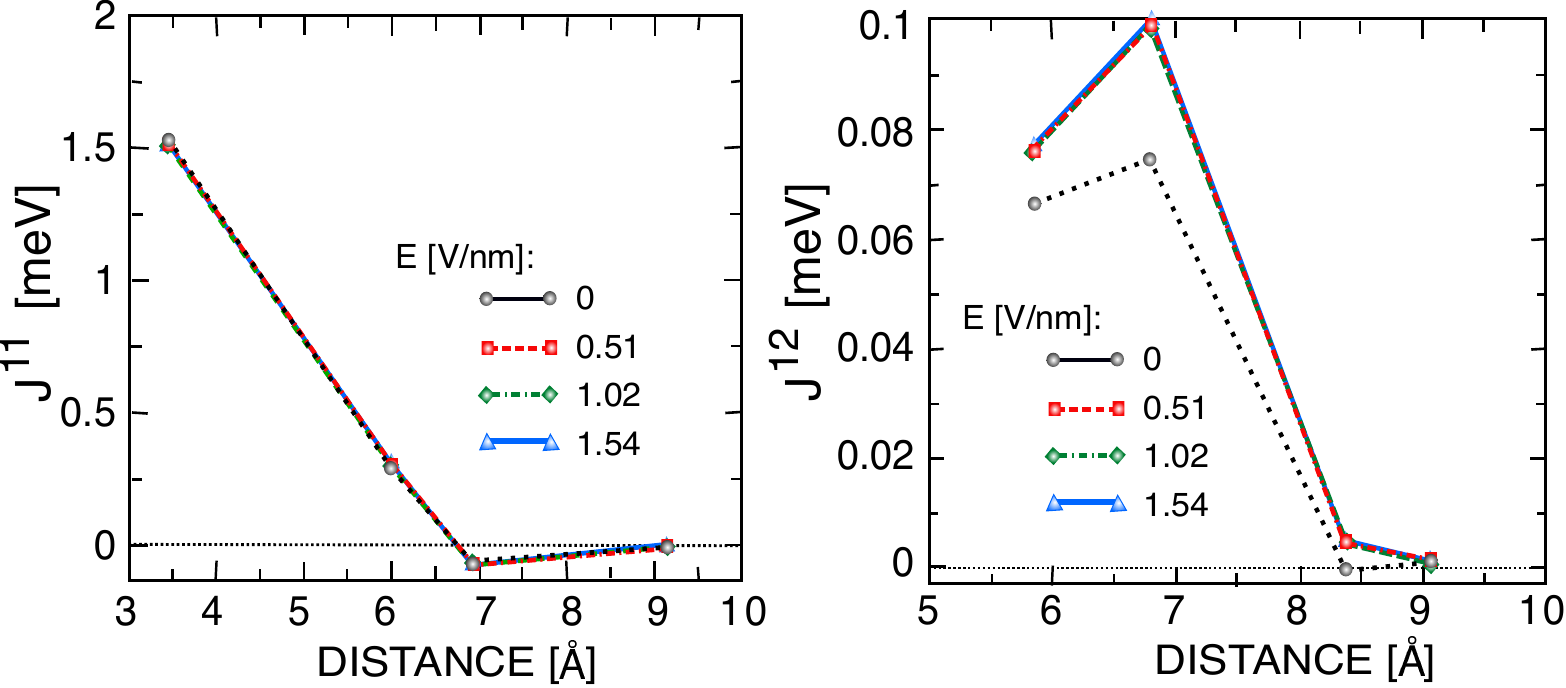} \\
\caption{(Color online) Total intralayer (left) and interlayer (right) exchange parameters of bilayer CrCl$_{3}$ as a function of electric field. A positive value indicates FM contribution, and a negative value indicates AFM contribution.
\label{fig:ps}}
\end{figure*}

Finally, we compute different components of the DMI in the unstrained CrCl$_{3}$ bilayer; see Fig.~\ref{fig:dm}. The amplitudes of the different components of the DMI vectors are small due to negligible SOC coupling in CrCl$_{3}$. Our calculations show that for both intra- and interlayer contributions, only NN and NNN DMIs can be nonzero. Contrary to the Heisenberg exchange interactions, the intra- and interlayer DMIs have similar orders of magnitude.
While the x- and z-components of the NN intralayer DMI vector, $D_x^{11}=D_x^{22}$ and $D_z^{11}=D_z^{22}$, are equal and finite, the y-component of the intralayer DMI vector is zero.
The situation is different for the interlayer DMI, where only the z-component of the NN DMI vector is nonzero.
The finite in-plane and out-of-plane DMI vectors are promising for engineering exotic textures and helicity-dependent magnonic phenomena in this bilayer.

\subsection{Effect of Biaxial Strain Field on Bilayer CrCl$_{3}$}
To explore the effect of strain fields on the spin interactions of bilayer CrCl$_{3}$, we apply an in-plane biaxial strain, defined as $\epsilon = (a - a_{0})/a_{0}$, where $a_{0}$ and $a$ are lattice parameters in the absence and presence of a strain field, respectively. Our calculations show that with the application of in-plane biaxial strain fields ranging from $\epsilon=-5\%$ to $\epsilon=+5\%$, the interlayer distance changes from $d=6.05 \AA$ to $d=5.78 \AA$. Additionally, the bond angle between the magnetic and nonmagnetic ions is changed by the strain field. In the following, we show how these changes affect different spin interactions.

\begin{figure*}[t]
\centering
\includegraphics[width=0.76 \textwidth]{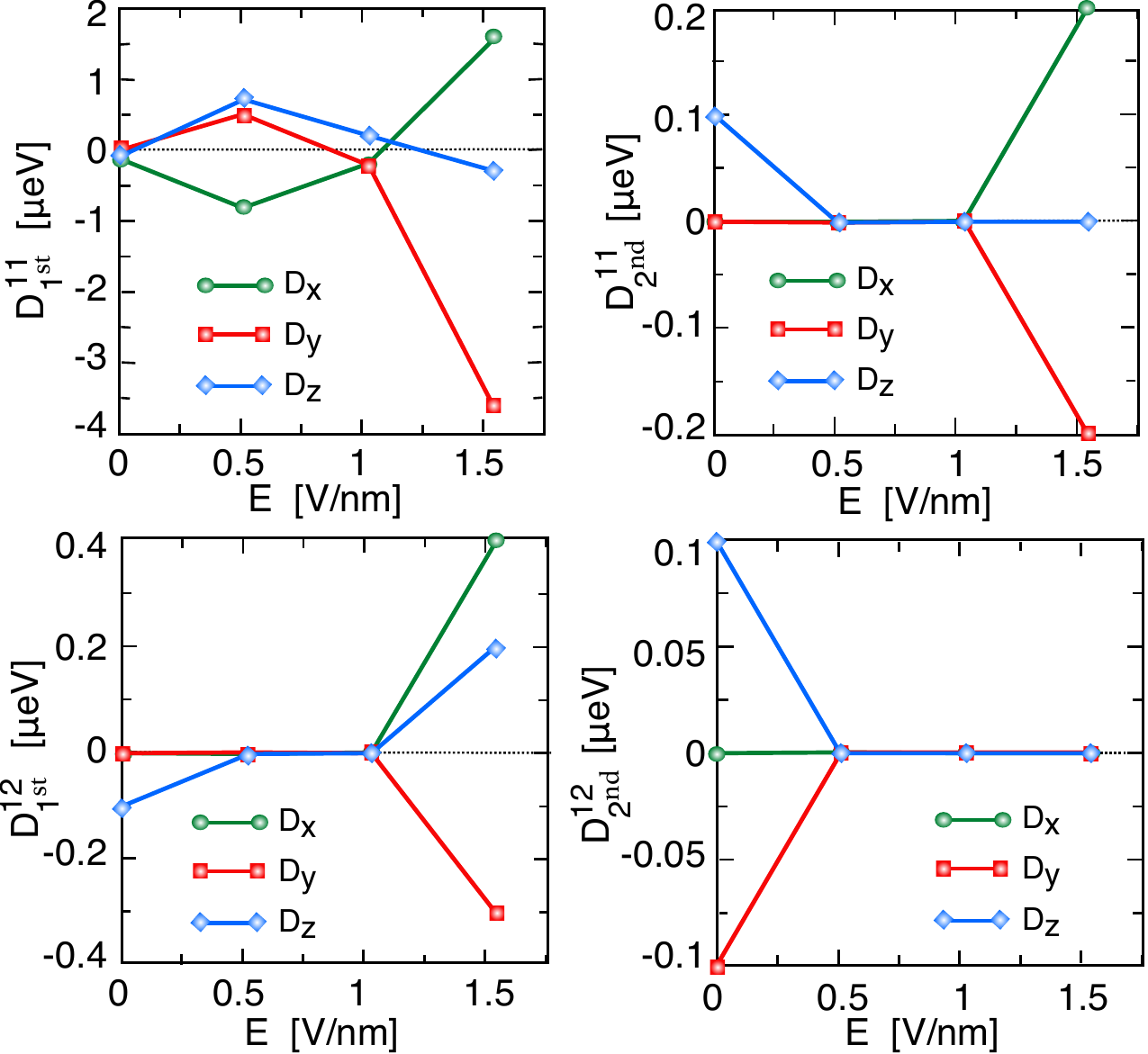} \\
\caption{(Color online) First- (left) and second-neighbor  (right) components of the intralayer (top panel) and interlayer (bottom panel) DMIs of bilayer CrCl$_{3}$ as a function of electric field. In the presence of an electric field, the $D_x$ and $D_y$ components of the NNN intralayer DMI increase, while $D_z$ decreases.
\label{fig:p2}}
\end{figure*}
First, we examine the magnetic ground state in the presence of various applied strain fields. Fig.~\ref{fig:Fig2} summarizes our ab initio calculation results.
By applying a tensile strain field $\epsilon>0$, both GGA and GGA+U calculations show that the total energy of the FM configuration is lower than that of the two other AFM configurations, and with increasing tensile strain, the FM configuration becomes more stable.

In contrast, bilayer CrCl$_{3}$ shows a quantum phase transition from the FM state to the AFM state when the compressive biaxial strain field ($\epsilon<0$) is larger than a threshold. Both GGA and GGA+U calculations show that with increasing compressive strain field, the G-type AFM state is more stable than the A-type AFM state. We obtain different critical compressive strain fields for GGA and GGA+U calculations, $\epsilon\approx -1\%$ and $\epsilon\approx -3\%$, respectively; see Fig.~\ref{fig:Fig2}. A similar magnetic phase transition was also reported for monolayer CrCl$_{3}$ at $\epsilon= -2.5\%$ compressive strain \cite{14}.

To investigate the origin of this phase transition, we compute bonding lengths and bonding angles in different configurations under various strain fields. The results are summarized in the Supplementary Information. 
Strain-induced changes in $\alpha_{\scriptscriptstyle{\mathrm{Cr-Cl-Cr}}}$ lead to changes in $t_{2g}$-$t_{2g}$ AFM and $e_{g}$-$t_{2g}$ FM couplings. Consequently, the phase transition from the FM to AFM state becomes possible.

We show the evolution of spin-spin interactions under the strain in Fig.~\ref{fig:F}. The values of intralayer and interlayer symmetric Heisenberg exchange coupling increase without changing the sign under tensile strains. As a result, the FM ground state remains unchanged with increasing lattice constant. Interestingly, our GGA+U calculations show that the sign of the NN intralayer symmetric Heisenberg exchange coupling is reversed above $3\%$ compressive strain. Fig. \ref{fig:F} shows that the sign of the NN intralayer symmetric exchange coefficient is negative at $-5\%$, confirming the phase transition from an FM phase to an AFM phase under compressive strain. The NN interlayer symmetric exchange becomes negligible at $5\%$ compressive strain due to the large interlayer distance.

The effects of strain fields on the intra- and interlayer DMIs are presented in Fig.~\ref{fig:dm}.
The results show that the NN and NNN intralayer components of the DMI vector increase when applying a tensile strain. As shown in Fig.~\ref{fig:dm}, upon $5\%$ tensile strain, the NN intralayers $D_z$ and $D_x$ become 10 times larger than the unstrained ones. Similarly, the NN and NNN interlayer components of the DMI vector increase with increasing tensile strain strength. Our ab initio calculations show that the NNN interlayers $D_{z}$ and $D_{y}$ are more than two times larger than the unstrained interlayers. Under $5\%$ compressive strain, the NN intralayers $D_x$ and $D_y$ become much larger than D$_z$, while the higher-nearest-neighbor intralayers are zero. In other words, the first-nearest-neighbor intralayer DMIs increase with decreasing intra-atomic distances, whereas the interlayer DMIs go to zero.

Furthermore, Fig.~\ref{fig:D2} shows that strain fields not only modify the strength of in-plane anisotropy but also change the direction of the magnetic anisotropy from in plane to out of plane. Our calculations show that by applying a tensile strain field, the in-plane magnetic anisotropy becomes negligible around $\approx+1\%$ and becomes out of plane for larger tensile strain fields. On the other hand, the in-plane anisotropy is increased by increasing the compressive strain fields.

 \subsection{Effect of Electric Field on Bilayer CrCl$_{3}$}
Electric field control of magnetic states and spin interactions is also an active area of research in spintronics due to its technological applications. It was shown previously that the magnetic and electric properties of the single layer of ${\mathrm{CrX_3}}$ can be controlled by gate voltages. In this part, we investigate the effect of perpendicular electric fields on a CrCl$_3$ bilayer. Our results show that within our ab initio methods, the energy difference between the AFM and FM states does not change much, and thus, the FM ground state remains the ground state of the system. Fig. \ref{fig:ps} shows that the applied electric field increases the NN and NNN interlayer exchange interactions and even makes the third NN interactions nonzero, while the intralayer exchange couplings remain nearly unchanged. These results are consistent with the robustness of the FM state based on our total energy analyses.

The effects of perpendicular electric fields on the magnitude and direction of the intralayer and interlayer DMI vectors are depicted in Fig.~\ref{fig:p2}. This figure shows that both the NN inter- and intralayer components of the DMI vector change the sign and that their amplitudes increase in the presence of a perpendicular electric field. These calculations show that at a small electric field, $D_z$ is the main component, whereas the in-plane components of the DMI vector have a dominant role at $E=1.5 eV$.
In the presence of an electric field, the magnitude of the in-plane component ($D_x$, $D_y$) of the NNN intralayer DMI increases, while the out-of-plane component ($D_z$) decreases. The electric field has an opposite effect on the NNN interlayer DMI, and they reduce to zero with increasing electric field.
Our calculations show that an electric field up to $1.52 V/nm$ cannot change the sign of the MAE of the CrCl$_{3}$ bilayer but increases the amplitude of the in-plane magnetic anisotropy.

\section{Summary and Concluding Remarks}\label{conclusion}
Efficient control of magnetic states and different spin interactions in a magnetic system is essential to the design of multifunctional spintronic nanodevices.
Recent theoretical and experimental studies on magnetic 2D vdW materials show the high flexibility of these systems. Our ab initio calculations show that bilayer CrCl${_3}$ is an interesting system. We show that this system can have a quantum magnetic phase transition from the FM state to the AFM state upon the application of a compressive strain field.  Thus, CrX3 2D systems with FM and AFM ground states form an ideal platform for investigating magnonic phenomena.

On the other hand, we show that the amplitude and direction of uniaxial magnetic anisotropy can be changed by strain fields. The thermal stability of a 2D magnetic system is directly related to a finite bandgap in its magnon spectra. By increasing the magnetic anisotropy, one can increase the magnon gap and the corresponding critical temperature. Tuning the critical temperature is important not only for designing room temperature nanodevices but also for achieving efficient and fast magnetic switching and studying the critical behavior of quantum transport properties close to the quantum and thermal phase transitions.

Finally, we find that the DMI vectors in this system can be tuned and modified by both strain and electric fields. Both the direction and amplitude of the DMI vectors are important for designing chiral magnetic states, topological magnetic textures, and topological magnonic states, as well as engineering spin-phonon interactions. We argue that using strain engineering in CrCl$_3$ one can observe and study exotic phenomena, such as the magnon spin Nernst effect in the AFM state \cite{Ran,Kovalev} and the realization of topological edge states in the FM state \cite{Yaruslav}, in one system.

Our ab initio calculations show that while symmetric exchange and uniaxial magnetic anisotropy are sensitive to strain fields and change with a perpendicular gate voltage, the DMI vectors can be tuned by both strain and electric fields.

\newpage
\section*{Data availability}
The datasets generated and/or analysed during the current study are available from the corresponding authors on reasonable request.

\section*{Acknowledgments}
This work was supported by the Norwegian Financial Mechanism 2014-2021 under the Polish-Norwegian Research Project NCN GRIEG '2Dtronics' No. 2019/34/H/ST3/00515. A.Q. was partially supported by the Research Council of Norway through its Centres of Excellence funding scheme, Project No. 262633, 'QuSpin'.

\clearpage
\newpage

\appendix
\section*{Supplementary Information}
\renewcommand\thefigure{\thesection.\arabic{figure}}
\setcounter{figure}{0}

\section{Robustness of Results Against Different U Values}
\label{app:U}
\begin{figure*}[h]
\centering
\includegraphics[width=0.480 \textwidth]{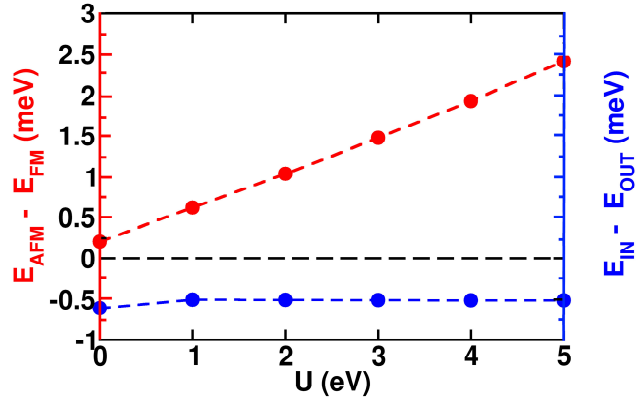} \\
\caption{(Color online) Energy differences between AFM and FM states and MAE under different Hubbard U parameters.}\label{fig:U}
\end{figure*}

Figure \ref{fig:U} shows the robustness of the magnetic ground-state against different Hubbard U values. In fact, the sign of magnetic anisotropy (${\mathrm{MAE}=E_{\scriptscriptstyle{\mathrm{IN}}}-E_{\scriptscriptstyle{\mathrm{OUT}}}}$) and the magnetic exchange interaction (E$_{\rm{AFM}}$ - E$_{\rm{FM}}$) are not changed as we change the onsite Hubbard parameter.

\section{Strain-induced Changes in the Bonding Angle and Bonding Length}
\label{app:Landalpha}
\begin{figure*}[h]
\centering
\includegraphics[width=0.850 \textwidth]{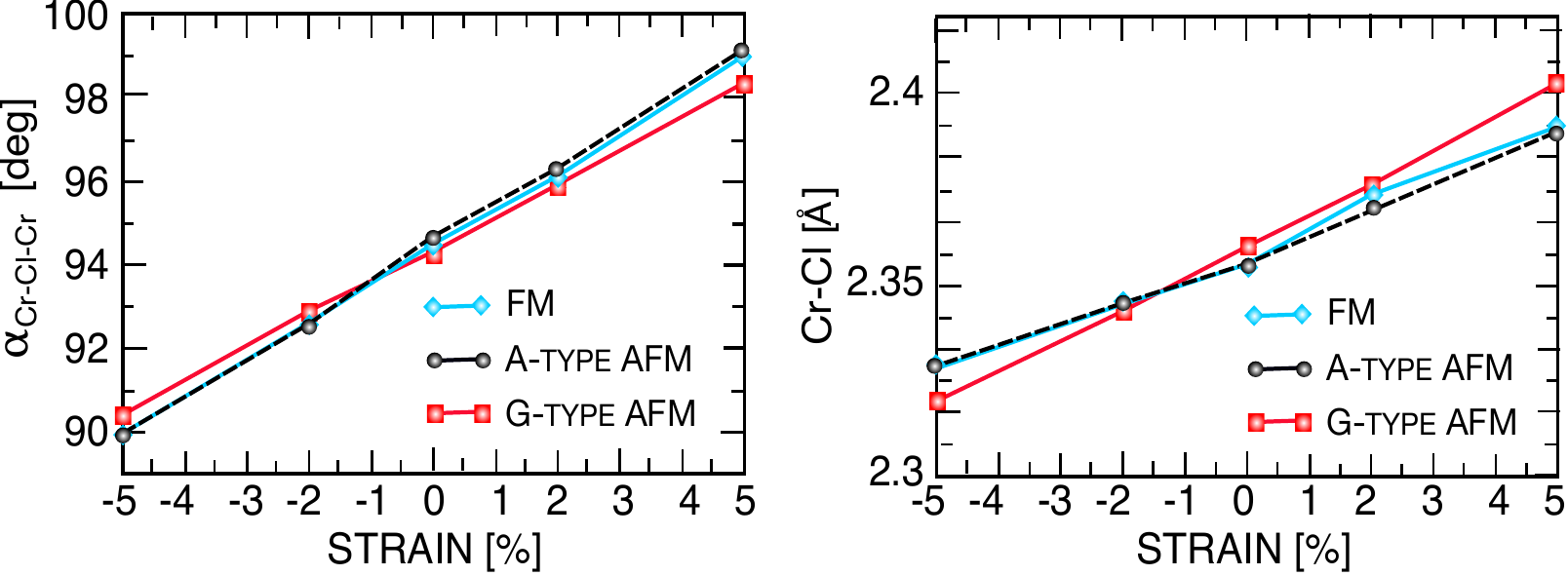} \\
\caption{(Color online) The Cr-Cl-Cr bonding angle ($\alpha_{Cr-Cl-Cr}$) (left) and the Cr-Cl bonding length (L$_{Cr-Cl}$) (right) in bilayer CrCl$_{3}$ under compressive (negative sign) and tensile (positive sign) biaxial strains. The $\alpha$ angle decreases with increasing compressive strain, while it increases toward a straight angle with increasing the tensile strain.}\label{fig:D}
\end{figure*}

The strain fields can alter the bonding angle, $\alpha_{\scriptscriptstyle{\mathrm{Cr-Cl-Cr}}}$, and bonding length, $L_{\scriptscriptstyle{\mathrm{Cr-Cl}}}$. For all FM, A-type AFM, and G-type AFM configurations, the tensile strain increases $\alpha_{\scriptscriptstyle{\mathrm{Cr-Cl-Cr}}}$, while this angle decreases under compressive strain. Figure \ref{fig:D} shows that the Cr-I bonding length slightly changes in the presence of strain fields.

\clearpage
\bibliography{bibfile}

\begin{thebibliography}{10}
\urlstyle{rm}
\expandafter\ifx\csname url\endcsname\relax
  \def\url#1{\texttt{#1}}\fi
\expandafter\ifx\csname urlprefix\endcsname\relax\def\urlprefix{URL }\fi
\expandafter\ifx\csname doiprefix\endcsname\relax\def\doiprefix{DOI: }\fi
\providecommand{\bibinfo}[2]{#2}
\providecommand{\eprint}[2][]{\url{#2}}

\bibitem{1}
\bibinfo{author}{Mounet, N.} \emph{et~al.}
\newblock \bibinfo{journal}{\bibinfo{title}{{Two-dimensional materials from
  high-throughput computational exfoliation of experimentally known
  compounds}}}.
\newblock {\emph{\JournalTitle{Nat. Nanotechnol.}}}
  \textbf{\bibinfo{volume}{13}}, \bibinfo{pages}{246--252},
  \doiprefix\url{10.1038/s41565-017-0035-5} (\bibinfo{year}{2018}).

\bibitem{2}
\bibinfo{author}{Feng, Y.~P.} \emph{et~al.}
\newblock \bibinfo{journal}{\bibinfo{title}{{Prospects of spintronics based on
  2D materials}}}.
\newblock {\emph{\JournalTitle{Wiley Interdiscip. Rev. Comput. Mol. Sci.}}}
  \textbf{\bibinfo{volume}{7}}, \bibinfo{pages}{e1313},
  \doiprefix\url{10.1002/wcms.1313} (\bibinfo{year}{2017}).

\bibitem{3}
\bibinfo{author}{Burch, K.~S.}, \bibinfo{author}{Mandrus, D.} \&
  \bibinfo{author}{Park, J.-G.}
\newblock \bibinfo{journal}{\bibinfo{title}{{Magnetism in two-dimensional van
  der Waals materials}}}.
\newblock {\emph{\JournalTitle{Nature}}} \textbf{\bibinfo{volume}{563}},
  \bibinfo{pages}{47--52}, \doiprefix\url{10.1038/s41586-018-0631-z}
  (\bibinfo{year}{2018}).

\bibitem{4}
\bibinfo{author}{Novoselov, K.}, \bibinfo{author}{Mishchenko},
  \bibinfo{author}{Carvalho, A.} \& \bibinfo{author}{Castro~Neto, A.}
\newblock \bibinfo{journal}{\bibinfo{title}{{2D materials and van der Waals
  heterostructures}}}.
\newblock {\emph{\JournalTitle{Science}}} \textbf{\bibinfo{volume}{353}},
  \bibinfo{pages}{6298}, \doiprefix\url{10.1126/science.aac9439}
  (\bibinfo{year}{2016}).

\bibitem{5}
\bibinfo{author}{Gibertini, M.}, \bibinfo{author}{Koperski, M.},
  \bibinfo{author}{Morpurgo, A.~F.} \& \bibinfo{author}{Novoselov, K.~S.}
\newblock \bibinfo{journal}{\bibinfo{title}{{Magnetic 2D materials and
  heterostructures}}}.
\newblock {\emph{\JournalTitle{Nat. Nanotechnol.}}}
  \textbf{\bibinfo{volume}{14}}, \bibinfo{pages}{408--419},
  \doiprefix\url{10.1038/s41565-019-0438-6} (\bibinfo{year}{2019}).

\bibitem{6}
\bibinfo{author}{Wang, M.-C.} \emph{et~al.}
\newblock \bibinfo{journal}{\bibinfo{title}{{Prospects and opportunities of 2D
  van der Waals magnetic systems}}}.
\newblock {\emph{\JournalTitle{Ann. Phys.}}} \textbf{\bibinfo{volume}{532}},
  \bibinfo{pages}{1900452}, \doiprefix\url{10.1002/andp.201900452}
  (\bibinfo{year}{2020}).

\bibitem{AlirezaUltrafast}
\bibinfo{author}{Losada, J.~M.}, \bibinfo{author}{Brataas, A.} \&
  \bibinfo{author}{Qaiumzadeh, A.}
\newblock \bibinfo{journal}{\bibinfo{title}{{Ultrafast control of spin
  interactions in honeycomb antiferromagnetic insulators}}}.
\newblock {\emph{\JournalTitle{Phys. Rev. B}}} \textbf{\bibinfo{volume}{100}},
  \bibinfo{pages}{060410}, \doiprefix\url{10.1103/PhysRevB.100.060410}
  (\bibinfo{year}{2019}).

\bibitem{AlirezaCrI3}
\bibinfo{author}{Vishkayi, S.~I.}, \bibinfo{author}{Torbatian, Z.},
  \bibinfo{author}{Qaiumzadeh, A.} \& \bibinfo{author}{Asgari, R.}
\newblock \bibinfo{journal}{\bibinfo{title}{{Strain and electric-field control
  of spin-spin interactions in monolayer {CrI}$_{3}$}}}.
\newblock {\emph{\JournalTitle{Phys. Rev. Mater.}}}
  \textbf{\bibinfo{volume}{4}}, \bibinfo{pages}{094004},
  \doiprefix\url{10.1103/PhysRevMaterials.4.094004} (\bibinfo{year}{2020}).

\bibitem{Banasree}
\bibinfo{author}{Sadhukhan, B.}, \bibinfo{author}{Bergman, A.},
  \bibinfo{author}{Kvashnin, Y.~O.}, \bibinfo{author}{Hellsvik, J.} \&
  \bibinfo{author}{Delin, A.}
\newblock \bibinfo{journal}{\bibinfo{title}{{Spin-lattice couplings in
  two-dimensional ${\mathrm{CrI}}_{3}$ from first-principles computations}}}.
\newblock {\emph{\JournalTitle{Phys. Rev. B}}} \textbf{\bibinfo{volume}{105}},
  \bibinfo{pages}{104418}, \doiprefix\url{10.1103/PhysRevB.105.104418}
  (\bibinfo{year}{2022}).

\bibitem{PhysRevLett.77.386}
\bibinfo{author}{Ivanov, B.~A.} \& \bibinfo{author}{Tartakovskaya, E.~V.}
\newblock \bibinfo{journal}{\bibinfo{title}{{Stabilization of Long-Range
  Magnetic Order in 2D Easy-Plane Antiferromagnets}}}.
\newblock {\emph{\JournalTitle{Phys. Rev. Lett.}}}
  \textbf{\bibinfo{volume}{77}}, \bibinfo{pages}{386--389},
  \doiprefix\url{10.1103/PhysRevLett.77.386} (\bibinfo{year}{1996}).

\bibitem{fridman2002stabilization}
\bibinfo{author}{Fridman, Y.~A.}, \bibinfo{author}{Spirin, D.},
  \bibinfo{author}{Alexeyev, C.} \& \bibinfo{author}{Matiunin, D.}
\newblock \bibinfo{journal}{\bibinfo{title}{{Stabilization of the long-range
  magnetic ordering by dipolar and magnetoelastic interactions in
  two-dimensional ferromagnets}}}.
\newblock {\emph{\JournalTitle{Eur. Phys. J. B}}}
  \textbf{\bibinfo{volume}{26}}, \bibinfo{pages}{185--190},
  \doiprefix\url{10.1140/epjb/e20020079} (\bibinfo{year}{2002}).

\bibitem{girovsky2017long}
\bibinfo{author}{Girovsky, J.} \emph{et~al.}
\newblock \bibinfo{journal}{\bibinfo{title}{{Long-range ferrimagnetic order in
  a two-dimensional supramolecular Kondo lattice}}}.
\newblock {\emph{\JournalTitle{Nat. Commun.}}} \textbf{\bibinfo{volume}{8}},
  \bibinfo{pages}{1--8}, \doiprefix\url{10.1038/ncomms15388}
  (\bibinfo{year}{2017}).

\bibitem{8}
\bibinfo{author}{Seyler, K.~L.} \emph{et~al.}
\newblock \bibinfo{journal}{\bibinfo{title}{{Ligand-field helical luminescence
  in a 2D ferromagnetic insulator}}}.
\newblock {\emph{\JournalTitle{Nat. Phys.}}} \textbf{\bibinfo{volume}{14}},
  \bibinfo{pages}{277--281}, \doiprefix\url{10.1038/s41567-017-0006-7}
  (\bibinfo{year}{2018}).

\bibitem{9}
\bibinfo{author}{Wang, Z.} \emph{et~al.}
\newblock \bibinfo{journal}{\bibinfo{title}{{Very large tunneling
  magnetoresistance in layered magnetic semiconductor CrI$_3$}}}.
\newblock {\emph{\JournalTitle{Nat. Commun.}}} \textbf{\bibinfo{volume}{9}},
  \bibinfo{pages}{1--8}, \doiprefix\url{10.1038/s41467-018-04953-8}
  (\bibinfo{year}{2018}).

\bibitem{10}
\bibinfo{author}{McGuire, M.~A.} \emph{et~al.}
\newblock \bibinfo{journal}{\bibinfo{title}{{Magnetic behavior and spin-lattice
  coupling in cleavable van der Waals layered CrCl$_3$ crystals}}}.
\newblock {\emph{\JournalTitle{Phys. Rev. Mater.}}}
  \textbf{\bibinfo{volume}{1}}, \bibinfo{pages}{014001},
  \doiprefix\url{10.1103/PhysRevMaterials.1.014001} (\bibinfo{year}{2017}).

\bibitem{11}
\bibinfo{author}{McGuire, M.~A.}, \bibinfo{author}{Dixit, H.},
  \bibinfo{author}{Cooper, V.~R.} \& \bibinfo{author}{Sales, B.~C.}
\newblock \bibinfo{journal}{\bibinfo{title}{{Coupling of crystal structure and
  magnetism in the layered, ferromagnetic insulator CrI$_3$}}}.
\newblock {\emph{\JournalTitle{Chem. Mater.}}} \textbf{\bibinfo{volume}{27}},
  \bibinfo{pages}{612--620}, \doiprefix\url{10.1021/cm504242t}
  (\bibinfo{year}{2015}).

\bibitem{song2021direct}
\bibinfo{author}{Song, T.} \emph{et~al.}
\newblock \bibinfo{journal}{\bibinfo{title}{{Direct visualization of magnetic
  domains and moir{\'e} magnetism in twisted 2D magnets}}}.
\newblock {\emph{\JournalTitle{Science}}} \textbf{\bibinfo{volume}{374}},
  \bibinfo{pages}{1140--1144}, \doiprefix\url{10.1126/science.abj7478}
  (\bibinfo{year}{2021}).

\bibitem{yao2021noncollinear}
\bibinfo{author}{Yao, X.}, \bibinfo{author}{Wang, Y.} \& \bibinfo{author}{Dong,
  S.}
\newblock \bibinfo{journal}{\bibinfo{title}{{Noncollinear topological textures
  in two-dimensional van der Waals materials: From magnetic to polar
  systems}}}.
\newblock {\emph{\JournalTitle{Int. J. Mod. Phys. B}}}
  \textbf{\bibinfo{volume}{35}}, \bibinfo{pages}{2130004},
  \doiprefix\url{10.1142/S0217979221300048} (\bibinfo{year}{2021}).

\bibitem{AlirezaDWDMI}
\bibinfo{author}{Qaiumzadeh, A.}, \bibinfo{author}{Kristiansen, L.~A.} \&
  \bibinfo{author}{Brataas, A.}
\newblock \bibinfo{journal}{\bibinfo{title}{{Controlling chiral domain walls in
  antiferromagnets using spin-wave helicity}}}.
\newblock {\emph{\JournalTitle{Phys. Rev. B}}} \textbf{\bibinfo{volume}{97}},
  \bibinfo{pages}{020402}, \doiprefix\url{10.1103/PhysRevB.97.020402}
  (\bibinfo{year}{2018}).

\bibitem{20}
\bibinfo{author}{Lei, C.} \emph{et~al.}
\newblock \bibinfo{journal}{\bibinfo{title}{{Magnetoelectric response of
  antiferromagnetic CrI$_3$ bilayers}}}.
\newblock {\emph{\JournalTitle{Nano Lett.}}} \textbf{\bibinfo{volume}{21}},
  \bibinfo{pages}{1948--1954}, \doiprefix\url{10.1021/acs.nanolett.0c04242}
  (\bibinfo{year}{2021}).

\bibitem{14}
\bibinfo{author}{Webster, L.} \& \bibinfo{author}{Yan, J.-A.}
\newblock \bibinfo{journal}{\bibinfo{title}{{Strain-tunable magnetic anisotropy
  in monolayer CrCl$_3$, CrBr$_3$, and CrI$_3$}}}.
\newblock {\emph{\JournalTitle{Phys. Rev. B}}} \textbf{\bibinfo{volume}{98}},
  \bibinfo{pages}{144411}, \doiprefix\url{10.1103/PhysRevB.98.144411}
  (\bibinfo{year}{2018}).

\bibitem{Song_2019}
\bibinfo{author}{Song, T.} \emph{et~al.}
\newblock \bibinfo{journal}{\bibinfo{title}{{Switching 2D magnetic states via
  pressure tuning of layer stacking}}}.
\newblock {\emph{\JournalTitle{Nat. Mater.}}} \textbf{\bibinfo{volume}{18}},
  \bibinfo{pages}{1298--1302}, \doiprefix\url{10.1038/s41563-019-0505-2}
  (\bibinfo{year}{2019}).

\bibitem{Li_2019}
\bibinfo{author}{Li, T.} \emph{et~al.}
\newblock \bibinfo{journal}{\bibinfo{title}{{Pressure-controlled interlayer
  magnetism in atomically thin {CrI}$_3$}}}.
\newblock {\emph{\JournalTitle{Nat. Mater.}}} \textbf{\bibinfo{volume}{18}},
  \bibinfo{pages}{1303--1308}, \doiprefix\url{10.1038/s41563-019-0506-1}
  (\bibinfo{year}{2019}).

\bibitem{12}
\bibinfo{author}{Huang, B.} \emph{et~al.}
\newblock \bibinfo{journal}{\bibinfo{title}{{Electrical control of 2D magnetism
  in bilayer CrI$_3$}}}.
\newblock {\emph{\JournalTitle{Nat. Nanotechnol.}}}
  \textbf{\bibinfo{volume}{13}}, \bibinfo{pages}{544--548},
  \doiprefix\url{10.1038/s41565-018-0121-3} (\bibinfo{year}{2018}).

\bibitem{13}
\bibinfo{author}{Jiang, S.}, \bibinfo{author}{Shan, J.} \&
  \bibinfo{author}{Mak, K.~F.}
\newblock \bibinfo{journal}{\bibinfo{title}{{Electric-field switching of
  two-dimensional van der Waals magnets}}}.
\newblock {\emph{\JournalTitle{Nat. Mater.}}} \textbf{\bibinfo{volume}{17}},
  \bibinfo{pages}{406--410}, \doiprefix\url{10.1038/s41563-018-0040-6}
  (\bibinfo{year}{2018}).

\bibitem{PhysRevLett.127.037204}
\bibinfo{author}{Dupont, M.} \emph{et~al.}
\newblock \bibinfo{journal}{\bibinfo{title}{{Monolayer ${\mathrm{CrCl}}_{3}$ as
  an Ideal Test Bed for the Universality Classes of 2D Magnetism}}}.
\newblock {\emph{\JournalTitle{Phys. Rev. Lett.}}}
  \textbf{\bibinfo{volume}{127}}, \bibinfo{pages}{037204},
  \doiprefix\url{10.1103/PhysRevLett.127.037204} (\bibinfo{year}{2021}).

\bibitem{Pizzochero_2020}
\bibinfo{author}{Pizzochero, M.} \& \bibinfo{author}{Yazyev, O.~V.}
\newblock \bibinfo{journal}{\bibinfo{title}{{Inducing Magnetic Phase
  Transitions in Monolayer CrI$_3$ via Lattice Deformations}}}.
\newblock {\emph{\JournalTitle{J. Phys. Chem. C}}}
  \textbf{\bibinfo{volume}{124}}, \bibinfo{pages}{7585--7590},
  \doiprefix\url{10.1021/acs.jpcc.0c01873} (\bibinfo{year}{2020}).

\bibitem{7}
\bibinfo{author}{Huang, B.} \emph{et~al.}
\newblock \bibinfo{journal}{\bibinfo{title}{{Layer-dependent ferromagnetism in
  a van der Waals crystal down to the monolayer limit}}}.
\newblock {\emph{\JournalTitle{Nature}}} \textbf{\bibinfo{volume}{546}},
  \bibinfo{pages}{270--273}, \doiprefix\url{10.1038/nature22391}
  (\bibinfo{year}{2017}).

\bibitem{pizzochero2020magnetic}
\bibinfo{author}{Pizzochero, M.}, \bibinfo{author}{Yadav, R.} \&
  \bibinfo{author}{Yazyev, O.~V.}
\newblock \bibinfo{journal}{\bibinfo{title}{{Magnetic exchange interactions in
  monolayer CrI3 from many-body wavefunction calculations}}}.
\newblock {\emph{\JournalTitle{2D Mater.}}} \textbf{\bibinfo{volume}{7}},
  \bibinfo{pages}{035005}, \doiprefix\url{10.1088/2053-1583/ab7cab}
  (\bibinfo{year}{2020}).

\bibitem{PhysRevB.94.075401}
\bibinfo{author}{Fransson, J.}, \bibinfo{author}{Black-Schaffer, A.~M.} \&
  \bibinfo{author}{Balatsky, A.~V.}
\newblock \bibinfo{journal}{\bibinfo{title}{Magnon dirac materials}}.
\newblock {\emph{\JournalTitle{Phys. Rev. B}}} \textbf{\bibinfo{volume}{94}},
  \bibinfo{pages}{075401}, \doiprefix\url{10.1103/PhysRevB.94.075401}
  (\bibinfo{year}{2016}).

\bibitem{olsen2021unified}
\bibinfo{author}{Olsen, T.}
\newblock \bibinfo{journal}{\bibinfo{title}{{Unified Treatment of Magnons and
  Excitons in Monolayer CrI$_3$ from Many-Body Perturbation Theory}}}.
\newblock {\emph{\JournalTitle{Phys. Rev. Lett.}}}
  \textbf{\bibinfo{volume}{127}}, \bibinfo{pages}{166402},
  \doiprefix\url{10.1103/PhysRevLett.127.166402} (\bibinfo{year}{2021}).

\bibitem{lee2020fundamental}
\bibinfo{author}{Lee, I.} \emph{et~al.}
\newblock \bibinfo{journal}{\bibinfo{title}{{Fundamental spin interactions
  underlying the magnetic anisotropy in the Kitaev ferromagnet CrI$_3$}}}.
\newblock {\emph{\JournalTitle{Phys. Rev. Lett.}}}
  \textbf{\bibinfo{volume}{124}}, \bibinfo{pages}{017201},
  \doiprefix\url{10.1103/PhysRevLett.124.017201} (\bibinfo{year}{2020}).

\bibitem{chen2018topological}
\bibinfo{author}{Chen, L.} \emph{et~al.}
\newblock \bibinfo{journal}{\bibinfo{title}{{Topological spin excitations in
  honeycomb ferromagnet CrI$_3$}}}.
\newblock {\emph{\JournalTitle{Phys. Rev. X}}} \textbf{\bibinfo{volume}{8}},
  \bibinfo{pages}{041028}, \doiprefix\url{10.1103/PhysRevX.8.041028}
  (\bibinfo{year}{2018}).

\bibitem{jaeschke2021theory}
\bibinfo{author}{Jaeschke-Ubiergo, R.}, \bibinfo{author}{Su\'arez~Morell, E.}
  \& \bibinfo{author}{Nunez, A.~S.}
\newblock \bibinfo{journal}{\bibinfo{title}{{Theory of magnetism in the van der
  Waals magnet ${\mathrm{CrI}}_{3}$}}}.
\newblock {\emph{\JournalTitle{Phys. Rev. B}}} \textbf{\bibinfo{volume}{103}},
  \bibinfo{pages}{174410}, \doiprefix\url{10.1103/PhysRevB.103.174410}
  (\bibinfo{year}{2021}).

\bibitem{ke2021electron}
\bibinfo{author}{Ke, L.} \& \bibinfo{author}{Katsnelson, M.~I.}
\newblock \bibinfo{journal}{\bibinfo{title}{{Electron correlation effects on
  exchange interactions and spin excitations in 2D van der Waals materials}}}.
\newblock {\emph{\JournalTitle{Npj Comput. Mater.}}}
  \textbf{\bibinfo{volume}{7}}, \bibinfo{pages}{1--8},
  \doiprefix\url{10.1038/s41524-020-00469-2} (\bibinfo{year}{2021}).

\bibitem{jiang2018electric}
\bibinfo{author}{Jiang, S.}, \bibinfo{author}{Shan, J.} \&
  \bibinfo{author}{Mak, K.~F.}
\newblock \bibinfo{journal}{\bibinfo{title}{{Electric-field switching of
  two-dimensional van der Waals magnets}}}.
\newblock {\emph{\JournalTitle{Nat. Mater.}}} \textbf{\bibinfo{volume}{17}},
  \bibinfo{pages}{406--410},
  \doiprefix\url{https://doi.org/10.1038/s41563-018-0040-6}
  (\bibinfo{year}{2018}).

\bibitem{morell2019control}
\bibinfo{author}{Morell, E.~S.}, \bibinfo{author}{Le{\'o}n, A.},
  \bibinfo{author}{Miwa, R.~H.} \& \bibinfo{author}{Vargas, P.}
\newblock \bibinfo{journal}{\bibinfo{title}{{Control of magnetism in bilayer
  CrI$_3$ by an external electric field}}}.
\newblock {\emph{\JournalTitle{2D Mater.}}} \textbf{\bibinfo{volume}{6}},
  \bibinfo{pages}{025020} (\bibinfo{year}{2019}).

\bibitem{15}
\bibinfo{author}{Liu, J.}, \bibinfo{author}{Mo, P.}, \bibinfo{author}{Shi, M.},
  \bibinfo{author}{Gao, D.} \& \bibinfo{author}{Lu, J.}
\newblock \bibinfo{journal}{\bibinfo{title}{{Multi-scale analysis of
  strain-dependent magnetocrystalline anisotropy and strain-induced Villari and
  Nagaoka-Honda effects in a two-dimensional ferromagnetic chromium tri-iodide
  monolayer}}}.
\newblock {\emph{\JournalTitle{J. Appl. Phys.}}}
  \textbf{\bibinfo{volume}{124}}, \bibinfo{pages}{044303},
  \doiprefix\url{10.1063/1.5036924} (\bibinfo{year}{2018}).

\bibitem{16}
\bibinfo{author}{Wu, Z.}, \bibinfo{author}{Yu, J.} \& \bibinfo{author}{Yuan,
  S.}
\newblock \bibinfo{journal}{\bibinfo{title}{{Strain-tunable magnetic and
  electronic properties of monolayer CrI$_3$}}}.
\newblock {\emph{\JournalTitle{Phys. Chem. Chem. Phys}}}
  \textbf{\bibinfo{volume}{21}}, \bibinfo{pages}{7750--7755},
  \doiprefix\url{10.1039/C8CP07067A} (\bibinfo{year}{2019}).

\bibitem{17}
\bibinfo{author}{Song, T.} \emph{et~al.}
\newblock \bibinfo{journal}{\bibinfo{title}{{Giant tunneling magnetoresistance
  in spin-filter van der Waals heterostructures}}}.
\newblock {\emph{\JournalTitle{Science}}} \textbf{\bibinfo{volume}{360}},
  \bibinfo{pages}{1214--1218}, \doiprefix\url{10.1126/science.aar4851}
  (\bibinfo{year}{2018}).

\bibitem{18}
\bibinfo{author}{Klein, D.~R.} \emph{et~al.}
\newblock \bibinfo{journal}{\bibinfo{title}{{Probing magnetism in 2D van der
  Waals crystalline insulators via electron tunneling}}}.
\newblock {\emph{\JournalTitle{Science}}} \textbf{\bibinfo{volume}{360}},
  \bibinfo{pages}{1218--1222}, \doiprefix\url{10.1126/science.aar3617}
  (\bibinfo{year}{2018}).

\bibitem{leon2020strain}
\bibinfo{author}{Le{\'o}n, A.}, \bibinfo{author}{Gonz{\'a}lez, J.},
  \bibinfo{author}{Mej{\'\i}a-L{\'o}pez, J.}, \bibinfo{author}{de~Lima, F.~C.}
  \& \bibinfo{author}{Morell, E.~S.}
\newblock \bibinfo{journal}{\bibinfo{title}{{Strain-induced phase transition in
  CrI$_3$ bilayers}}}.
\newblock {\emph{\JournalTitle{2D Mater.}}} \textbf{\bibinfo{volume}{7}},
  \bibinfo{pages}{035008} (\bibinfo{year}{2020}).

\bibitem{19}
\bibinfo{author}{Lado, J.~L.} \& \bibinfo{author}{Fern{\'a}ndez-Rossier, J.}
\newblock \bibinfo{journal}{\bibinfo{title}{{On the origin of magnetic
  anisotropy in two dimensional CrI$_3$}}}.
\newblock {\emph{\JournalTitle{2D Mater.}}} \textbf{\bibinfo{volume}{4}},
  \bibinfo{pages}{035002}, \doiprefix\url{10.1088/2053-1583/aa75ed}
  (\bibinfo{year}{2017}).

\bibitem{21}
\bibinfo{author}{Soriano, D.}, \bibinfo{author}{Cardoso, C.} \&
  \bibinfo{author}{Fern{\'a}ndez-Rossier, J.}
\newblock \bibinfo{journal}{\bibinfo{title}{{Interplay between interlayer
  exchange and stacking in CrI$_3$ bilayers}}}.
\newblock {\emph{\JournalTitle{Solid State Commun.}}}
  \textbf{\bibinfo{volume}{299}}, \bibinfo{pages}{113662},
  \doiprefix\url{10.1016/j.ssc.2019.113662} (\bibinfo{year}{2019}).

\bibitem{22}
\bibinfo{author}{Jiang, P.} \emph{et~al.}
\newblock \bibinfo{journal}{\bibinfo{title}{{Stacking tunable interlayer
  magnetism in bilayer CrI$_3$}}}.
\newblock {\emph{\JournalTitle{Phys. Rev. B}}} \textbf{\bibinfo{volume}{99}},
  \bibinfo{pages}{144401}, \doiprefix\url{10.1103/PhysRevB.99.144401}
  (\bibinfo{year}{2019}).

\bibitem{bed}
\bibinfo{author}{Bedoya-Pinto, A.} \emph{et~al.}
\newblock \bibinfo{journal}{\bibinfo{title}{{Intrinsic 2D-XY ferromagnetism in
  a van der Waals monolayer}}}.
\newblock {\emph{\JournalTitle{Science}}} \textbf{\bibinfo{volume}{374}},
  \bibinfo{pages}{616--620}, \doiprefix\url{10.1126/science.abd5146}
  (\bibinfo{year}{2021}).

\bibitem{32}
\bibinfo{author}{Klein, D.~R.} \emph{et~al.}
\newblock \bibinfo{journal}{\bibinfo{title}{{Enhancement of interlayer exchange
  in an ultrathin two-dimensional magnet}}}.
\newblock {\emph{\JournalTitle{Nat. Phys.}}} \textbf{\bibinfo{volume}{15}},
  \bibinfo{pages}{1255--1260}, \doiprefix\url{10.1038/s41567-019-0651-0}
  (\bibinfo{year}{2019}).

\bibitem{25}
\bibinfo{author}{Giannozzi, P.} \emph{et~al.}
\newblock \bibinfo{journal}{\bibinfo{title}{{QUANTUM ESPRESSO: a modular and
  open-source software project for quantum simulations of materials}}}.
\newblock {\emph{\JournalTitle{J. Phys. Condens. Matter}}}
  \textbf{\bibinfo{volume}{21}}, \bibinfo{pages}{395502},
  \doiprefix\url{10.1088/0953-8984/21/39/395502} (\bibinfo{year}{2009}).

\bibitem{26}
\bibinfo{author}{Perdew, J.~P.}, \bibinfo{author}{Burke, K.} \&
  \bibinfo{author}{Ernzerhof, M.}
\newblock \bibinfo{journal}{\bibinfo{title}{{Generalized gradient approximation
  made simple}}}.
\newblock {\emph{\JournalTitle{Phys. Rev. Lett.}}}
  \textbf{\bibinfo{volume}{77}}, \bibinfo{pages}{3865},
  \doiprefix\url{10.1103/PhysRevLett.77.3865} (\bibinfo{year}{1996}).

\bibitem{29}
\bibinfo{author}{Grimme, S.}
\newblock \bibinfo{journal}{\bibinfo{title}{{Semiempirical GGA-type density
  functional constructed with a long-range dispersion correction}}}.
\newblock {\emph{\JournalTitle{J. Comput. Chem.}}}
  \textbf{\bibinfo{volume}{27}}, \bibinfo{pages}{1787--1799},
  \doiprefix\url{10.1002/jcc.20495} (\bibinfo{year}{2006}).

\bibitem{30}
\bibinfo{author}{Cococcioni, M.} \& \bibinfo{author}{De~Gironcoli, S.}
\newblock \bibinfo{journal}{\bibinfo{title}{{Linear response approach to the
  calculation of the effective interaction parameters in the LDA+U method}}}.
\newblock {\emph{\JournalTitle{Phys. Rev. B}}} \textbf{\bibinfo{volume}{71}},
  \bibinfo{pages}{035105}, \doiprefix\url{10.1103/PhysRevB.71.035105}
  (\bibinfo{year}{2005}).

\bibitem{jang2019}
\bibinfo{author}{Jang, S.~W.}, \bibinfo{author}{Jeong, M.~Y.},
  \bibinfo{author}{Yoon, H.}, \bibinfo{author}{Ryee, S.} \&
  \bibinfo{author}{Han, M.~J.}
\newblock \bibinfo{journal}{\bibinfo{title}{{Microscopic understanding of
  magnetic interactions in bilayer ${\mathrm{CrI}}_{3}$}}}.
\newblock {\emph{\JournalTitle{Phys. Rev. Materials}}}
  \textbf{\bibinfo{volume}{3}}, \bibinfo{pages}{031001},
  \doiprefix\url{10.1103/PhysRevMaterials.3.031001} (\bibinfo{year}{2019}).

\bibitem{27}
\bibinfo{author}{Li, D.}, \bibinfo{author}{Barreteau, C.},
  \bibinfo{author}{Castell, M.~R.}, \bibinfo{author}{Silly, F.} \&
  \bibinfo{author}{Smogunov, A.}
\newblock \bibinfo{journal}{\bibinfo{title}{{Out-versus in-plane magnetic
  anisotropy of free Fe and Co nanocrystals: Tight-binding and first-principles
  studies}}}.
\newblock {\emph{\JournalTitle{Phys. Rev. B}}} \textbf{\bibinfo{volume}{90}},
  \bibinfo{pages}{205409}, \doiprefix\url{10.1103/PhysRevB.90.205409}
  (\bibinfo{year}{2014}).

\bibitem{28}
\bibinfo{author}{Mackintosh, A.~R.} \& \bibinfo{author}{Andersen, O.~K.}
\newblock \bibinfo{title}{{The electronic structure of transition metals}}.
\newblock In \bibinfo{editor}{Springford, M.} (ed.)
  \emph{\bibinfo{booktitle}{Electrons at the Fermi Surface}},
  chap.~\bibinfo{chapter}{5} (\bibinfo{publisher}{Cambridge University Press,
  Cambridge, England}, \bibinfo{year}{1980}).

\bibitem{31}
\bibinfo{author}{He, X.}, \bibinfo{author}{Helbig, N.},
  \bibinfo{author}{Verstraete, M.~J.} \& \bibinfo{author}{Bousquet, E.}
\newblock \bibinfo{journal}{\bibinfo{title}{{TB2J: A python package for
  computing magnetic interaction parameters}}}.
\newblock {\emph{\JournalTitle{Comput. Phys. Commun.}}}
  \textbf{\bibinfo{volume}{264}}, \bibinfo{pages}{107938},
  \doiprefix\url{10.1016/j.cpc.2021.107938} (\bibinfo{year}{2021}).

\bibitem{AlirezaDMI1}
\bibinfo{author}{Ado, I.~A.}, \bibinfo{author}{Qaiumzadeh, A.},
  \bibinfo{author}{Brataas, A.} \& \bibinfo{author}{Titov, M.}
\newblock \bibinfo{journal}{\bibinfo{title}{{Chiral ferromagnetism beyond
  Lifshitz invariants}}}.
\newblock {\emph{\JournalTitle{Phys. Rev. B}}} \textbf{\bibinfo{volume}{101}},
  \bibinfo{pages}{161403}, \doiprefix\url{10.1103/PhysRevB.101.161403}
  (\bibinfo{year}{2020}).

\bibitem{AlirezaDMI2}
\bibinfo{author}{Qaiumzadeh, A.}, \bibinfo{author}{Ado, I.~A.},
  \bibinfo{author}{Duine, R.~A.}, \bibinfo{author}{Titov, M.} \&
  \bibinfo{author}{Brataas, A.}
\newblock \bibinfo{journal}{\bibinfo{title}{{Theory of the Interfacial
  Dzyaloshinskii-Moriya Interaction in Rashba Antiferromagnets}}}.
\newblock {\emph{\JournalTitle{Phys. Rev. Lett.}}}
  \textbf{\bibinfo{volume}{120}}, \bibinfo{pages}{197202},
  \doiprefix\url{10.1103/PhysRevLett.120.197202} (\bibinfo{year}{2018}).

\bibitem{AlirezaDMI3}
\bibinfo{author}{Ado, I.~A.}, \bibinfo{author}{Qaiumzadeh, A.},
  \bibinfo{author}{Duine, R.~A.}, \bibinfo{author}{Brataas, A.} \&
  \bibinfo{author}{Titov, M.}
\newblock \bibinfo{journal}{\bibinfo{title}{{Asymmetric and Symmetric Exchange
  in a Generalized 2D Rashba Ferromagnet}}}.
\newblock {\emph{\JournalTitle{Phys. Rev. Lett.}}}
  \textbf{\bibinfo{volume}{121}}, \bibinfo{pages}{086802},
  \doiprefix\url{10.1103/PhysRevLett.121.086802} (\bibinfo{year}{2018}).

\bibitem{LIECHTENSTEIN198765}
\bibinfo{author}{Liechtenstein, A.}, \bibinfo{author}{Katsnelson, M.},
  \bibinfo{author}{Antropov, V.} \& \bibinfo{author}{Gubanov, V.}
\newblock \bibinfo{journal}{\bibinfo{title}{Local spin density functional
  approach to the theory of exchange interactions in ferromagnetic metals and
  alloys}}.
\newblock {\emph{\JournalTitle{Journal of Magnetism and Magnetic Materials}}}
  \textbf{\bibinfo{volume}{67}}, \bibinfo{pages}{65--74},
  \doiprefix\url{https://doi.org/10.1016/0304-8853(87)90721-9}
  (\bibinfo{year}{1987}).

\bibitem{Ran}
\bibinfo{author}{Cheng, R.}, \bibinfo{author}{Okamoto, S.} \&
  \bibinfo{author}{Xiao, D.}
\newblock \bibinfo{journal}{\bibinfo{title}{{Spin Nernst Effect of Magnons in
  Collinear Antiferromagnets}}}.
\newblock {\emph{\JournalTitle{Phys. Rev. Lett.}}}
  \textbf{\bibinfo{volume}{117}}, \bibinfo{pages}{217202},
  \doiprefix\url{10.1103/PhysRevLett.117.217202} (\bibinfo{year}{2016}).

\bibitem{Kovalev}
\bibinfo{author}{Zyuzin, V.~A.} \& \bibinfo{author}{Kovalev, A.~A.}
\newblock \bibinfo{journal}{\bibinfo{title}{{Magnon Spin Nernst Effect in
  Antiferromagnets}}}.
\newblock {\emph{\JournalTitle{Phys. Rev. Lett.}}}
  \textbf{\bibinfo{volume}{117}}, \bibinfo{pages}{217203},
  \doiprefix\url{10.1103/PhysRevLett.117.217203} (\bibinfo{year}{2016}).

\bibitem{Yaruslav}
\bibinfo{author}{Kim, S.~K.}, \bibinfo{author}{Ochoa, H.},
  \bibinfo{author}{Zarzuela, R.} \& \bibinfo{author}{Tserkovnyak, Y.}
\newblock \bibinfo{journal}{\bibinfo{title}{{Realization of the
  Haldane-Kane-Mele Model in a System of Localized Spins}}}.
\newblock {\emph{\JournalTitle{Phys. Rev. Lett.}}}
  \textbf{\bibinfo{volume}{117}}, \bibinfo{pages}{227201},
  \doiprefix\url{10.1103/PhysRevLett.117.227201} (\bibinfo{year}{2016}).

\end{thebibliography}

\end{document}